\documentclass[pra,twocolumn,showpacs,preprintnumbers,amsmath,amssymb]{revtex4-1}

\usepackage{graphicx}
\usepackage{dcolumn}
\usepackage{bm}

\begin{document}
\title{The two-level atom laser: analytical results and the laser transition.}

\author{Paul Gartner}
\affiliation{Institute for Theoretical Physics,
        University of Bremen, 28334 Bremen, Germany \\
         and \\
         National Institute of Materials Physics,
        Bucharest-M\u agurele, Romania}

\email{gartner@itp.uni-bremen.de}

\date{\today}

\begin{abstract}
The problem of the two-level atom laser is studied analytically. 
The steady-state solution is expressed as a continued fraction, and
allows for accurate approximation by rational functions. Moreover, we
show that the abrupt change observed in the pump dependence of the
steady-state population is directly connected with the transition to
the lasing regime. The condition for a sharp transition to 
Poissonian statistics is expressed as a scaling limit of vanishing cavity
loss and light-matter coupling, $\kappa \to 0$, $g \to 0$,
such that $g^2/\kappa$ stays finite and $g^2/\kappa > 2 \gamma$, where
$\gamma$ is the rate of atomic losses. The same scaling procedure is
also shown to describe a similar change to Poisson distribution in
the Scully-Lamb laser model too, suggesting that the low-$\kappa$,
low-$g$ asymptotics is of a more general significance for the laser
transition.   
\end{abstract}


\maketitle

\section{Introduction}

A single two-level emitter in interaction with a cavity mode is the 
simplest illustration of cavity quantum electrodynamics. Actually, in
experimental situations the emitter, either a true atom or an
artificial one (quantum dot), usually brings into the picture more
than just two levels. Besides the pair of states which couple with the
laser mode, other states are involved in the carrier excitation and
scattering processes. It was early realized \cite{mu-sav} that
incoherent pumping directly from the lasing levels destroys the
coherence between these and the photon output is strongly reduced for
large pump rates. This quenching
effect acts against lasing and can be avoided by involving additional
states. More recent realizations of single-emitter lasers
\cite{kim,arak-opex} were modelled using at least four levels
\cite{booz-kim, arak-nat}. In a semiconductor quantum dot (QD) one
typically encounters many-level, many-carrier situations which give
rise to multiexcitonic effects. Moreover, in the presence of a spectral
continuum (e.g. wetting-layer extended states) where, in typical
experimental situations carriers are initially created by optical or
electrical pumping, the creation of fermionic reservoirs becomes
possible, from which carriers are injected into the QD
\cite{ben-yam}. The same charge distributions in the continuum are
also a source of screening and additional dephasing
\cite{lrke_06}. All these effects are clearly beyond the reach of a
two-level description.   

Despite this, the two-level emitter model still enjoys a 
wide popularity due to its simplicity and transparence and
remains a subject of intense investigations  
\cite{delv-fer, auff}. 
Also, models with more levels are sometimes mapped into
effective two-level ones \cite{pedr,booz}. Many
aspects of the light-matter interaction, like spectral properties
associated with strong coupling, are quite well described in this
framework \cite{delv-fer,auff,loeff,moerk, cui, delv-mol}.       

An appealing feature of the two-level model is the fact that the carrier
degrees of freedom are easily eliminated \cite{agwl} leading
to closed equations for the purely photonic statistics. This is a
feature shared with the Scully-Lamb random injection model
\cite{scl-lamb}, which addresses directly the photonic density
matrix. Therefore these cases are particularly appropriate for the
study of the laser transition, and indeed, the Scully-Lamb model became
a paradigm and a textbook case in this respect 
\cite{deg, sten, walls, orsz}. In comparison, with a few exceptions
\cite{karl, kil}, the two-level emitter has not enjoyed the same
attention from this point of view.   

In the Scully-Lamb model it is shown that the transition to lasing
corresponds to the peak of the photon number distribution getting
detached from the origin  and moving away to larger and larger
values. A thermal, exponential distribution changes into a Poissonian
statistics. In terms of the Glauber-Sudarshan
representation \cite{carm}, this means that the so-called
quasi-distribution function, initially located around the origin of
the complex plane, becomes narrowly concentrated on a ring of
increasing radius. Such a behaviour was also obtained numerically  
\cite{mu-sav, ginz} or by using analytic approximations
\cite{carm}, in several different contexts.    
    
In the case of the two-level model, the numerical simulations show an
abrupt change of regime, most clearly seen on the steady-state population
behaviour: the strictly concave dependence of the upper-level
population on the pumping rate becomes with a high accuracy linear, 
in the interval of intermediate pump rates \cite{delv-mol, pedr, karl}. 
Various numerical evidence suggest that this transition is associated
with the onset of lasing, but the connection to the behaviour of the
Glauber-Sudarshan function, like in the random injection model, is far
from obvious.   

In the present paper we address the two-level laser problem by
analytic methods. We show that the steady-state solution, usually
obtained by the long-time limit of the Liouville-von Neumann evolution,
can be expressed directly as a continued fraction involving the model
parameters. Exact relations as well as simple, yet accurate 
approximations for the solution can be derived in some cases from the
full formula. The theory of continued fractions also provides error
control for these approximations.  
 
But what is more important, the method also allows for a detailed
characterization of the laser transition. The appearance of a jump in
the solution (or some derivative) has its analog in the problem of
phase transitions, to which lasing was sometimes compared, and the
role of exactly soluble models is to clarify the mechanism of the
jump.

The two-level atom turns out to be such an exactly soluble, yet
nontrivial example on which the laser transition can be studied.
The condition for the appearance of the above-mentioned abrupt change
in the pump dependence of the population is proven to be equivalent to
the vanishing of the Glauber-Sudarshan function at the origin.
Moreover, one can show that this vanishing requirement is met only in
a certain scaling limit for the model parameters, and then it also
entails a Poissonian photon statistics.

Similarities with the thermodynamic limit in phase transitions
and appropriate scaling asymptotics, like the Grad limit or the
weak-coupling, long-time limit, in the theory of markovian kinetics
\cite{spohn} come to mind. In all these examples a regime change
occurs, rigorously speaking, only in a certain asymptotic parameter
domain, and this is precisely what takes place in our case too. Of
course, with parameters close but slightly away from the scaling limit
the transition may still be seen, albeit less sharply.
 
The paper is organized as follows: 
After defining the model (Sec.~\ref{sec:model}) 
the three-term recursion equation obeyed by an infinite sequence of
photon expectation values \cite{agwl} is discussed and its solution
expressed as a continued fraction. It is shown that  
{\em outside the interval of linear behaviour} the continued fraction
converges rapidly and its truncation  provides simple and accurate
approximants (Sec.~\ref{sec:cont_frac}). 
The discussion of the linear regime is more complicated and requires
an extension of the recursion relation. This is achieved by noting
that the photonic expectation values are moments of the Glauber-Sudarshan 
function, and it is shown that the linear behaviour is equivalent to
the condition that this function vanishes at the origin (Sec.~\ref{sec:GS}). 
 
The next sections deal with the circumstances in which this condition
is met. Heuristic suggestions in this sense are obtained using the
rate-equation approximation (Sec.~\ref{sec:rateq}). Turning back to
the full problem (Sec.~\ref{sec:scale}) it is shown that exact
vanishing occurs as a limiting case, in which the rate of cavity
losses $\kappa $ and the light-matter coupling strength $g$ go
simultaneously to zero so that $g^2/\kappa = const$. A similar scaling
limit was obtained by Rice and Carmichael, as a condition for a sharp
lasing transition in the rate-equation approach to the many-emitter
laser \cite{rice-carm}. It should be noted that in contradistinction
to this many-emitter result, in the present case, as the pump increases
the system undergoes two transitions, a first one at the onset of
lasing and a second one at larger pump values, when the lasing is
destroyed by quenching. It is also shown that in this limiting
procedure, in the lasing regime the Glauber-Sudarshan function not
only vanishes at the origin but also becomes $\delta$-like distributed
on a ring centered at the origin, in agreement with the expected
Poissonian photon statistics. The radius of this ring starts
increasing from zero when the pump reaches the threshold value but
eventually decreases back to zero due to quenching.     

The conclusions regarding the scaling limit and the laser
transition go beyond the particular case of the two-level laser, as
checked on the Scully-Lamb model (see Appendix~\ref{sec:rand_inj}). 
   
\section{The model and its steady-state behaviour}
\label{sec:model}

We consider a two-level emitter (atom or QD) in
Jaynes-Cummings (JC) interaction with a cavity mode. Cavity losses,
spontaneous recombination into non-lasing modes and incoherent pumping
are included in the density operator evolution through Lindblad
terms with rates $\kappa$, $\gamma$ and $P$, respectively. 
Assuming perfect resonance between the emitter and the cavity
mode one has, in the interaction representation, the following
Liouville-von Neumann equation (in $\hbar = 1$ units) 
 
\begin{eqnarray}
 \frac{\partial}{\partial t} \rho =-i \left[ H_{JC},\rho \right]  &+&
    \frac{\kappa}{2} \left[2 b \,\rho\, b^{\dagger} - b^{\dagger}b \,\rho -
      \rho\, b^{\dagger}b\right] \nonumber \\
    &+& \frac{\gamma}{2} \left[2v^{\dagger}c\, \rho\, c^{\dagger}v -
      c^{\dagger}c\, \rho - \rho \, c^{\dagger}c \right]   \nonumber \\
    &+& \frac{P}{2} \left[2c^{\dagger}v\, \rho\, v^{\dagger}c -
      v^{\dagger}v\, \rho - \rho \, v^{\dagger}v \right] \; .
\label{eq:LvN}
\end{eqnarray}
The corresponding interaction Hamiltonian is given by
\begin{equation}
H_{JC} = g \left[b^{\dagger}v^{\dagger}c + b c^{\dagger}v\right] \; .
\label{equ:JC} 
\end{equation}
The creation and annihilation operators are $b^{\dagger}, b$ for the
laser mode, $c^{\dagger}, c$ for the upper (conduction band) level and
$v^{\dagger}, v$ for the lower (valence band) one. Only one carrier is
present in the system, so that one has $c^{\dagger}c+v^{\dagger}v=1$.   
We use QD notations throughout, the atomic pseudospin formalism being
easily recovered via $\sigma=v^{\dagger}c$,
$\sigma^{\dagger}=c^{\dagger}v$,
$\sigma^{\dagger}\sigma=c^{\dagger}c$, a.s.o.

The Liouville-von Neumann equation can be solved numerically until the
steady-state solution is reached. Then various steady-state
expectation values, regarding level populations and photon statistics
can be obtained as a function of the pump  rate $P$ or other
parameters \cite{delv-fer,auff, pedr}.

\begin{figure*}[!htb]
\begin{center}
\resizebox{0.85\textwidth}{!}{%
 \includegraphics{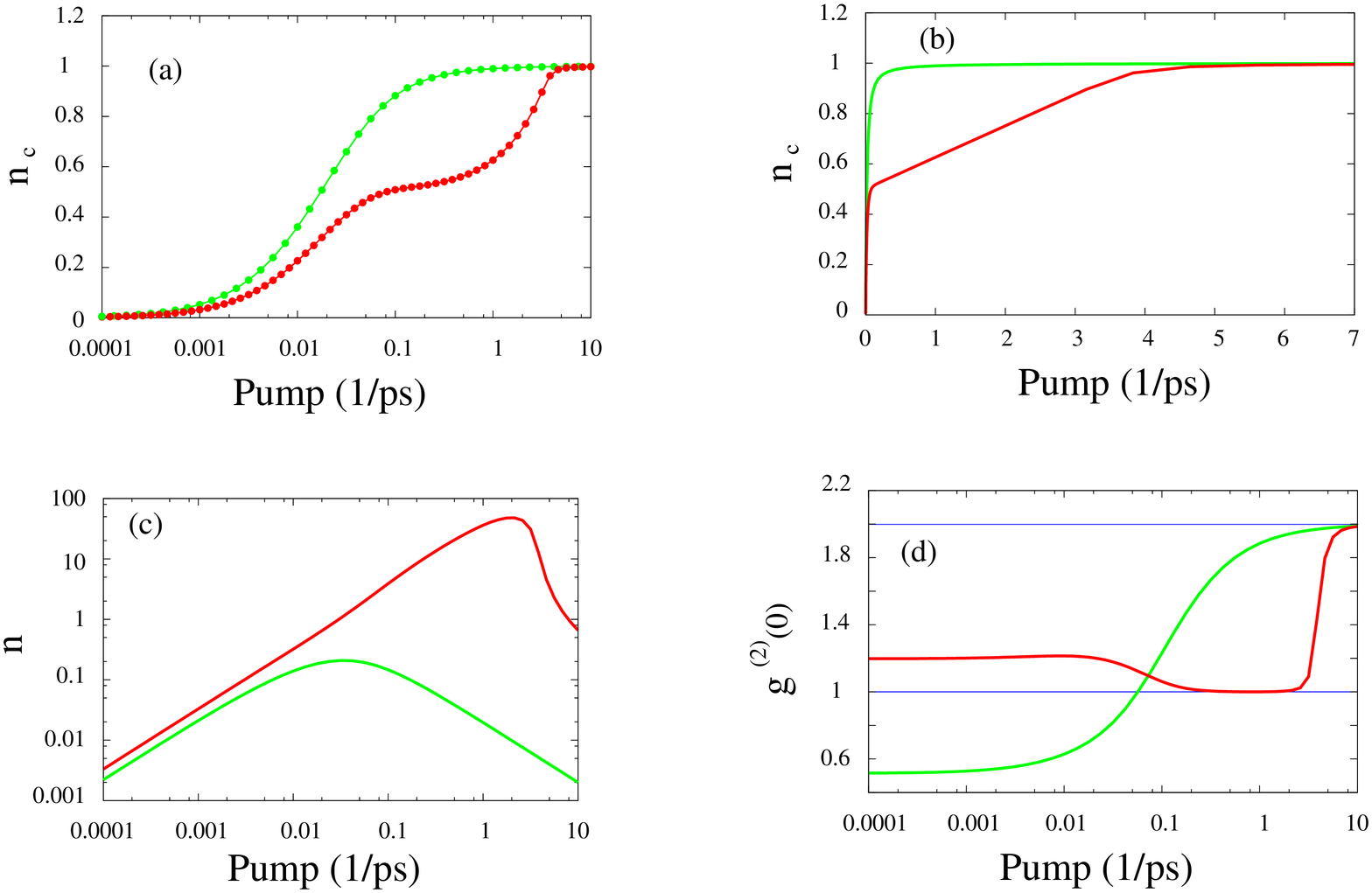} } 
\caption {\label{fig:Fig1} Steady state solutions of the
  Liouville-von Neumann equation vs. pump rate $P$. The parameters are:
Set A (red lines): $\kappa = 0.01/ps,\, g = 0.1/ps, \, \gamma =
0.02/ps$ and Set B (green lines) : $\kappa = 0.02/ps,\, g = 0.01/ps,
\, \gamma = 0.01/ps$. (a) the 
upper state population $n_c$, semi-log plot. Circles
correspond to the solution obtained by continued fraction (see
Sec~\ref{sec:cont_frac}). (b) the same in linear plot, showing the
appearance of the linear middle regime. (c) average photon number $n$.
(d) second-order autocorrelation function.} 
\end{center}
\end{figure*} 

Examining families of such plots \cite{delv-fer} one clearly detects
two types of behaviour. One example of each of these types is
shown in Fig.1 (see the Caption for the parameters used). The two
cases are best distinguished by their pump dependence of the
upper-level population $n_c=\left<c^{\dagger}c\right>$. In a
semi-logarithmic plot one may have, depending on the parameters,
either a simple S-shaped curve or a double S-shaped one (Fig.1a). The
difference is more striking in a linear plot (Fig.1b), where in the
first case one has a strictly concave function of $P$, while in the
second there appears an almost perfectly linear shortcut in the middle
$P$-region. This is strongly reminiscent of the behaviour of
thermodynamic potentials in the presence of a phase transition
\cite{stan}. 

In the present case too, the transition from concave to linear
dependence in the middle $P$ interval is a clear simptom of an
abrupt regime change in the functioning of the two-level laser. 
The figure shows that this ``linear middle'' regime (as it will be
subsequently called) is associated with inversion of population ($n_c
> 1/2$), a large photon output $n=\left<b^{\dagger}b\right>$
(Fig.1c), and a value of the second-order photon autocorrelation
function at zero delay $g^{(2)}(0) =
\left< b^{\dagger}b^{\dagger}bb\right>/\left<b^{\dagger}b\right>^2$
close to unity (Fig.1d). All these are indicative of the onset of
lasing which, depending on the parameters, may take place or not, as
illustrated by the two types of behaviour of Fig.1.  

In what follows it will be proven that the linear pump
dependence of the population is indeed associated with 
a Poissonian photon statistics, characteristic for coherent light. 
At the same time, the conditions for the presence or absence of the
transition, as well as cases in which the transition
becomes sharp will be identified.   

\section{Solution by continued fractions}
\label{sec:cont_frac} 

It is easy to see that the Liouville-von Neumann equation leads to a
closed system of equations for a subset of density matrix elements. In
the basis of QD-states $\left| c \right>,\left| v \right> $ and photon
number $n$, these are the 
diagonal ones $\rho^{cc}_{n,n}$, $\rho^{vv}_{n,n}$ and only  
those off-diagonal elements between states having the same number of
excitations $c^{\dagger}c+b^{\dagger}b$, i.e. $\rho^{cv}_{n,n+1}$ and
their complex conjugates $\rho^{vc}_{n+1,n}$. The latter restriction
is a consequence of the JC Hamiltonian, which conserves the excitation
number.       

Equivalently, one may limit oneself to the following set of
expectation values
\begin{eqnarray}
 c_k &=& \left < c^{\dagger}c \,  b^{\dagger\,k}b^k \right> \; ,\nonumber \\ 
 v_k &=& \left < v^{\dagger}v \, b^{\dagger\,k}b^k \right> \;, \qquad
 \qquad  k = 0,1,2 \ldots \quad \text{and} \nonumber \\ 
 t_k &=& -i g\left < v^{\dagger}c\,  b^{\dagger\,k}b^{k-1} \right> \; ,
 \quad \,\, k = 1,2,3 \ldots
\label{eq:cvt}
\end{eqnarray}
Obviously, one has $n_c=\left< c^{\dagger}c\right> =c_0$ and
$n_v=\left< v^{\dagger}v\right> =v_0$, and $c_0+v_0=1$. Of
particular interest are the purely photonic expectation values 
\begin{equation}
  p_k = c_k+v_k = \left< b^{\dagger\,k}b^k \right> \; ,
\label{eq:pk}
\end{equation}
and especially the low-index ones $p_1 = n$, and $p_2$ related to the
autocorrelation function by $p_2=g^{(2)}(0)\;p_1^2$. 
 
With the imaginary unit as a prefactor, the multi-photon-assisted
polarization $t_k$ is real. From Eq.~(\ref{eq:LvN}) the equations of
motion for these expectation values can be derived, together with the
steady-state conditions
\begin{subequations} 
\label{eq:eom}
\begin{eqnarray}
\frac{\partial}{\partial t} c_k &=&
     -(k\kappa+\gamma)c_k + P v_k -2 t_{k+1} = 0 \;  ,\label{eq:eoma}\\
\frac{\partial}{\partial t} v_k &=&
     \gamma c_k -(k\kappa+P)v_k +2t_{k+1} +2kt_k = 0 \; , \label{eq:eomb}\\
\frac{\partial}{\partial t} t_k &=&
      g^2c_k + g^2 kc_{k-1} -g^2 v_k  \nonumber \\
      && \hspace{1.5cm} -\,\frac{(2k-1)\kappa+P+\gamma}{2}t_k = 0 \; .    
      \label{eq:eomc} 
\end{eqnarray}
\end{subequations}   
By adding Eqs.~(\ref{eq:eoma}) and ~(\ref{eq:eomb}) one obtains the 
steady-state balance relation between the cavity losses and its feeding
through the photon-assisted polarization 
\begin{equation}
  \kappa \; p_k = 2 \; t_k \; , \quad k\geqslant 1 \; . \label{eq:balance} 
\end{equation}
Using this condition, Eq.~(\ref{eq:eoma}) for $k=0$ can be written as
$P\,n_v=\gamma\, n_c + \kappa\,n$, which leads to
\begin{equation}
n_c = \frac{P-\kappa\,n}{P+\gamma}  \; . \label{eq:n_c}
\end{equation}
This shows that the knowledge of the photon output $n$ is sufficient
for the determination of the level occupancies.    

More generally, from Eqs.~(\ref{eq:eoma}) and ~(\ref{eq:eomb}) for
arbitrary $k$, the unknowns $c_k$ and $v_k$ can be eliminated in favour of
$t_k$, and using again the balance condition Eq.~(\ref{eq:balance}) one
obtains closed equations for the photonic expectation values $p_k$, in
the form of a three-term recursion relation, first obtained by Agarwal
and Dutta Gupta \cite{agwl}  
\begin{equation}  
 A_k\,p_{k+1} + B_k\, p_k - C_k \, p_{k-1} = 0 \; , \quad k \geqslant
 1 \; , \label{eq:recrel} 
\end{equation}
with
\begin{eqnarray}
A_k &=& \frac{2\,\kappa}{k\, \kappa + P + \gamma} \; ,\nonumber \\
B_k &=& \frac{k\, \kappa -P+ \gamma}{k\, \kappa + P + \gamma}  +
        \frac{k\,\kappa}{(k-1)\,\kappa + P + \gamma} \nonumber \\ 
    && \hspace{2cm}  +\, \kappa \, \frac{(2k-1)\, \kappa + P
          +\gamma}{4g^2} \; ,\nonumber \\ 
C_k &=& \frac{k\,P}{(k-1)\,\kappa + P + \gamma} \; . \label{eq:ABC}  
\end{eqnarray}
This is an important feature of the model, allowing direct access to
the photon statistics after the elimination of the information
regarding the atom subsystem. 

Solving Eq.~(\ref{eq:recrel}) recursively requires knowledge of two
initial conditions, and the difficult point is that only one is
available, namely $p_0=1$. On the other hand, it is known
\cite{cont-fr} that the three-term recursion problem can be addressed
using continued fraction techniques \cite{agwl,cir}. To
this end Eq.~(\ref{eq:recrel}) is rewritten as
\begin{equation} 
  r_k + \lambda_k - \frac{\mu_k}{r_{k-1}} =0  \; , \quad k \geqslant
 1 \; , \label{eq:r}
\end{equation}  
with $r_k = p_{k+1}/p_k\,$, $\lambda_k = B_k/A_k\,$ and $\mu_k =
C_k/A_k\,$.
 
As before, the initial condition $r_0 = n$ is not known. On the other
hand it turns out that the large $k$ limit is easy to obtain. Indeed,
according to their definitions all $r_k$ should be positive and from
Eq.~(\ref{eq:r}) the positivity requirement for both $r_{k-1}$ and
$r_k$ leads to
\begin{equation}
0 \; \leqslant \;r_{k-1} \; \leqslant \; \frac{\mu_k}{\lambda_k} \; = \;
\frac{C_k}{B_k}\; .
\end{equation}   
An examination of Eq.~(\ref{eq:ABC}) shows that the upper bound goes
to zero as $k \to \infty$, and therefore $\lim_{k\to\infty}r_k=0$.  

With this 'initial' condition at $k=\infty$,
Eq.~(\ref{eq:r}) is solved by a backward propagation of the recursion
relation, $r_{k-1}=\mu_k/(\lambda_k+r_k)$.  This leads to a continued
fraction expansion of the solution, in terms of the parameters $\mu_k$
and $\lambda_k$  
\begin{equation}
\begin{split}
r_0 = \cfrac{\mu_1}{\lambda_1+r_1}  
    &= \cfrac{\mu_1}{\lambda_1+\cfrac{\mu_2}{\lambda_2 +r_2}} 
    = \; \ldots \; \\
    &= \cfrac{\mu_1}{\lambda_1+\cfrac{\mu_2}{\lambda_2
     +\cfrac{\mu_3}{\lambda_3+ \ldots}}} \; .
\end{split}
\label{eq:cont_fr}
\end{equation}
 
The continued fraction is convergent \cite{fn-1} and
the result is numerically in perfect agreement with the long time
limit of the Liouville-von Neumann solution (see Fig.1a). 

Approximate algebraic expressions $r_0^{(k)}$ for the solution $r_0$
are obtained by setting in Eq.~(\ref{eq:cont_fr}) $r_k=0$. Fig.2a shows a
comparison of the full solution $n=r_0$ with its second approximant
$r_0^{(2)}$ which is a rational function of $P$ whose expression
follows straightforwardly from Eq.~(\ref{eq:ABC}) and
$r_0^{(2)}=C_1B_2/(B_1B_2+A_1C_2)$.    
The corresponding approximants for $n_c$ obtained by
Eq.~(\ref{eq:n_c}) are shown in linear (Fig.2b) and semi-log (Fig.2c) plots
respectively. The agreement is remarkable on the whole range of pump
values in the case when the transition is absent (Fig.2d), and when
the transition does take place, the agreement remains good for the
pump values away from the linear middle interval. 

A closer inspection of the numerics shows that inside this interval
the continued fraction becomes much more slowly convergent. The
reason for this is the fact that the first values of 
$B_k$, and with them those of $\lambda_k$, are negative, resulting in
negative values for the low-ranking approximants of the positive
quantity $r_0$. Indeed, from Eq.~(\ref{eq:ABC}) it is clear
that $B_1$ may take negative values due to its first term. For this to
happen $P$ should not be too small. On the other hand for large values
of $P$ the last term becomes dominant, making $B_1$ positive again. In
other words, $B_1$ can be negative but only for an appropriate choice
of the parameters $\kappa, \gamma$ and $g$, and only for
intermediate pump values. This makes the analysis of the linear middle
interval look rather complicated.  

\begin{figure*}[!htb]
\begin{center}
\resizebox{0.85\textwidth}{!}{%
 \includegraphics{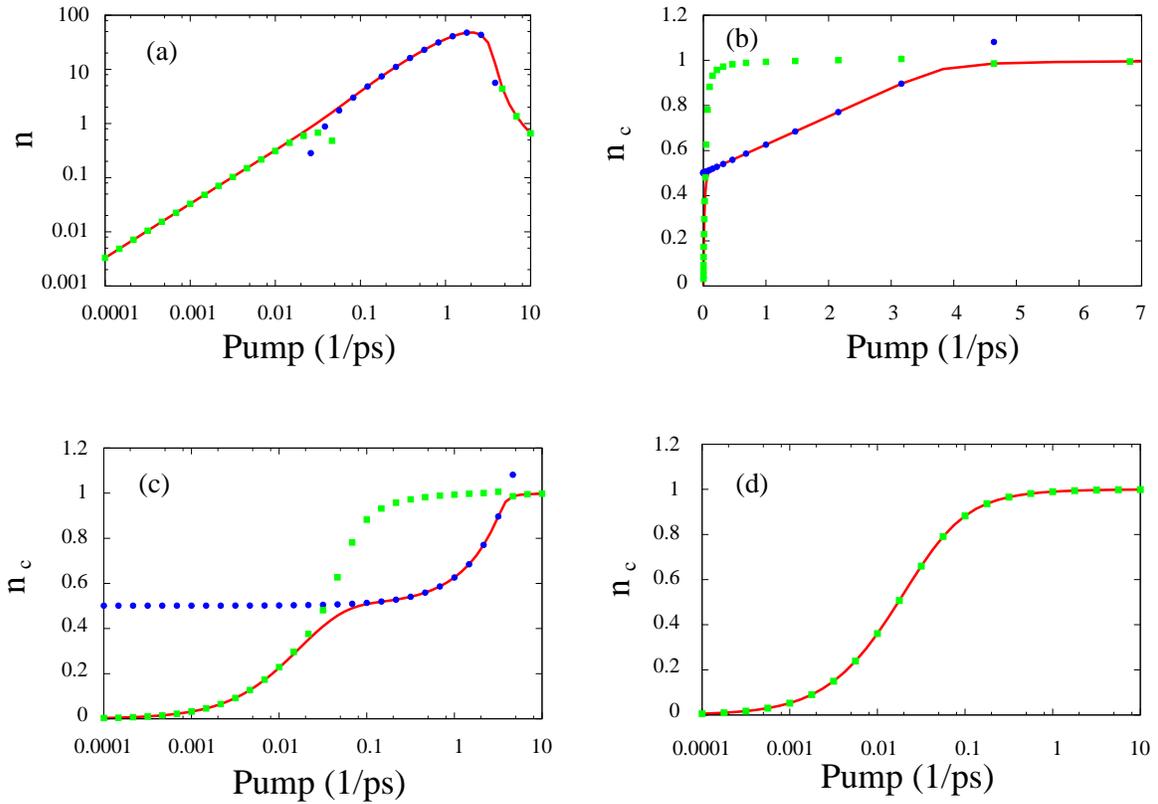}} 
\caption {\label{fig:Fig2} Comparison between the full solution (red
  line), the second approximant to the continued fraction (green
  squares) and the linear middle ansatz,
  Eqs.~(\ref{eq:ansatz_n}-\ref{eq:ansatz_cc}) (blue
  circles). The parameters are the same as in Fig.1: Set A for panels
  (a), (b), (c), and Set B for panel (d).  
}
\end{center}
\end{figure*} 

Surprisingly, the numerical evidence shows that the situation is quite
simple. One remarks that the full solution obeys to a good accuracy the
approximation provided by the ansatz
\begin{equation}   
  r_0 = -\lambda_0 = - \frac{B_0}{A_0} = 
                      \frac{P-\gamma}{2\kappa} -
                      \frac{(P+\gamma)(P+\gamma-\kappa)}{8g^2} \; ,
\label{eq:ansatz_n}
\end{equation} 
which leads, using Eq.~(\ref{eq:n_c}), to a linear expression for the
upper-state population     
\begin{equation} 
n_c = \frac{1}{2}+\frac{\kappa(P+\gamma-\kappa)}{8g^2} \; .
\label{eq:ansatz_cc}
\end{equation}
Moreover, in terms of the recursion relation, this ansatz is equivalent to the
assumption that Eq.~(\ref{eq:recrel}) would hold {\em even for $k=0$}
with the last term  $C_0 p_{-1}$ supposedly vanishing. Such an
interpretation was put forward by del Valle and Laussy 
\cite{delv-mol}, based on the argument that $C_0=0$. 
But this definitely requires a more careful 
analysis, because on the one hand $C_0$ comes multiplied by the
ill-defined $p_{-1}$ and on the other hand the ansatz is obviously
not working for all parameter values.     

Nevertheless, it is clear (see Fig.2) that the values given by
Eqs.~(\ref{eq:ansatz_n}) and ~(\ref{eq:ansatz_cc}) describe successfully
the linear middle regime. Moreover, since this regime is supposedly
connected with lasing, it is important to clarify the deeper
meaning of the ansatz. In other words one has to answer two basic
questions:  
(i) if indeed Eq.~(\ref{eq:recrel}) can be extended for $k=0$, which is
the right interpretation of its last term in this case and 
(ii) why is this term vanishing for parameters corresponding to the
linear middle regime.     

The first question is answered in the next Section and the second in
Section \ref{sec:scale}. 

Before closing this Section two remarks are in order. First, it is
easy to see that if $B_1$ is positive, all $B_k$ are positive too, and 
then the continued fraction coefficients are positive. In this case
\cite{cont-fr} the odd approximants $r_0^{(2k+1)}$ form a sequence of
decreasing upper bounds for the true solution $r_0$, while the even
approximants $r_0^{(2k)}$ give rise to an increasing sequence of lower bounds. 
This provides an efficient control of the truncation error, when the
system is away from the linear middle regime.
   
Second, we show that the
continued fraction solution provides exact analytic expressions for
the autocorrelation function for both the low- and the large-pump
limit. One has
\begin{equation} 
 g^{(2)}(0) = \frac{p_2}{p_1^2} = \frac{r_1}{r_0} \; ,
\label{eq:g2}
\end{equation}
with $r_0$ as in Eq.~(\ref{eq:cont_fr})  and similarly $r_1$ given by   
\begin{equation} 
 r_1 = {\cfrac{\mu_2}{\lambda_2+\cfrac{\mu_3}{\lambda_3 +\ldots}}} \; .
\label{eq:cont_fr_1}
\end{equation} 
As $P \to 0$ all $\mu_k$ vanish linearly while $\lambda_k$
stay finite. Therefore, in this limit one has
\begin{equation} 
 g^{(2)}(0) = \frac{\mu_2\;\lambda_1}{\mu_1\; \lambda_2} 
           =  \frac{C_2\;B_1}{C_1\; B_2} \; ,
\label{eq:g2_asympt}
\end{equation}
and as a result one obtains      
\begin{equation} 
 g^{(2)}(0)|_{P=0} = 2\;\frac{\kappa+\gamma}{3\kappa+\gamma} 
                   \;\frac{4g^2+\kappa\gamma}{4g^2+\kappa(\kappa+\gamma)} \; .
\label{eq:g2_P0}
\end{equation}
In the large $P$ asymptotics $\mu_k \sim P k / 2\kappa $ grow linearly with
$P$ and therefore are dominated by the quadratically increasing
$\lambda_k\sim P^2/8g^2$. In this case too the asymptotic behaviour is
given by Eq.~(\ref{eq:g2_asympt}) with the outcome $\lim_{P \to
  \infty}g^{(2)}(0)=2$. These facts are borne out by the numerical
results.   

\section{Glauber-Sudarshan representation}
\label{sec:GS} 

An alternative approach to the photon statistics is based on the
expansion of operators in the photonic Hilbert space as 
integrals over coherent-state projectors \cite{carm}. This
is known as the Glauber-Sudarshan (GS) or phase-space representation.
 
In our case the Hilbert space is enlarged by the QD degrees of
freedom and the density operator can be represented as a $2 \times2$
matrix made of blocks acting on the photonic degrees of freedom. For
each of these blocks the GS expansion can be used 
\begin{equation}
 \rho = \left( \begin{array}{cc}
               \rho^{cc}& \rho_{cv} \\ \rho^{cv}& \rho^{vv}
               \end{array} \right)  
      = \int \frac{{\text d}\alpha}{\pi} \left| \alpha \right>
         \left( \begin{array}{cc}
                R_c & \alpha Q \\ \alpha^* Q^*& R_v 
                \end{array} \right) \left< \alpha \right| \; .
\label{eq:GS}
\end{equation}
The GS functions $R_c$ and $R_v$ representing $\rho^{cc}$ and respectively
$\rho^{vv}$, are in principle functions of the complex argument
$\alpha$. Nevertheless, as discussed in Sec.~(\ref{sec:cont_frac}) the
elements of the density matrix which are diagonal in the QD-level
index are diagonal in the photon number too, and this entails that
the corresponding GS functions depend only on the radial parameter $s =
|\alpha|^2$ and are angle-independent $R_c = R_c(s)$, $R_v =
R_v(s)$. Similarly, the off-diagonal block $\rho^{cv}$ has non-zero
entries only one step above the diagonal. This corresponds to a GS
function with the structure $\alpha Q(s)$. Also, in accordance with
Eq.~(\ref{eq:cvt}) it is convenient to introduce $T(s)= -igQ(s)$, which
will turn out to be real.         

The advantage of using the GS representation becomes obvious after
expressing the expectation values of Eq.~(\ref{eq:cvt}) in this
formalism. Indeed, the procedure is convenient for calculating
averages of normal-ordered photonic operators 
\cite{orsz, carm}. One obtains
\begin{equation}
\begin{split}
 c_k &= \int_0^{\infty} R_c(s) s^k {\text d}s  \; , \quad 
 v_k = \int_0^{\infty} R_v(s) s^k {\text d}s  \; , \\ 
 t_k &= \int_0^{\infty} T(s) s^k {\text d}s    \; .
\end{split}
\label{eq:cvt_mom}
\end{equation}
This shows that the quantities of interest are moments of the GS
functions, and it is clear that this allows to extend their definition. 
For instance, now it is possible to define $t_0$ according to
Eq.~(\ref{eq:cvt_mom}), whereas in Eq.~(\ref{eq:cvt}) it made no sense.   

The purely photonic expectation values are obtained, as in
Eq.~(\ref{eq:pk}), by tracing over the QD-indices
\begin{equation}
p_k = c_k+v_k =  \int_0^{\infty} R(s) s^k {\text d}s  \; ,
\label{eq:pk_mom}
\end{equation}   
in which $R(s) = R_c(s)+R_v(s)$. 
For further reference let us remind here that the photon number
probability distribution $\rho_{n,n}=\rho^{cc}_{n,n}+\rho^{vv}_{n,n}$
is given in terms of $R(s)$ by
\begin{equation}
 \rho_{n,n} =  \int_0^{\infty} R(s) \frac{s^n}{n!}\, e^{-s} {\text d}s  \; .
\label{eq:rho_n_gs}
\end{equation}  
The Poissonian statistics is generated by a $\delta$-like GS function,
$R(s)=\delta(s-s_0)$, which corresponds in terms of the photonic expectation
values $p_k$ to an exponential behaviour $p_k=s_0^k$. 

The translation of the Liouville-von Neumann Eq.~(\ref{eq:LvN}) into the
phase-space language leads to Fokker-Planck equations for the GS functions
\cite{carm}, obtained by the correspondence  $b \to
\alpha$ and $b^{\dagger} \to \alpha^*-\partial/\partial \alpha$ when
the photonic operators act on $\rho$ from the left and
$b \to \alpha-\partial/\partial \alpha^*$ and $b^{\dagger} \to
\alpha^*$ when they act from the right. Taking into account that the
GS functions appearing here depend only on the radial argument, one
obtains the counterpart of  Eqs.~(\ref{eq:eom}) in the form  
\begin{subequations} 
\label{eq:fkpk}
\begin{eqnarray}
\frac{\partial}{\partial t} R_c &=&
     -2sT+\kappa (R_c+sR'_c)  \nonumber \\
     && \hspace{1.5cm} -\, \gamma R_c+PR_v = 0 \;  ,\label{eq:fkpka}\\
\frac{\partial}{\partial t} R_v &=&
  -2sT' +2(s-1)T + \kappa (R_v+sR'_v) \nonumber \\
     && \hspace{1.5cm} +\, \gamma R_c -PR_v = 0\; , \label{eq:fkpkb}\\
\frac{\partial}{\partial t} T &=&
      g^2(R_c-R_v) -g^2R'_c + \kappa (\frac{3}{2}T+sT')  \nonumber \\
     && \hspace{1.5cm} -\, \frac{\gamma +P}{2}T = 0 \; ,    \label{eq:fkpkc}
\end{eqnarray}
\end{subequations}   
where the prime stands for the derivative with respect to $s$.
By adding Eqs.~(\ref{eq:fkpka}) and ~(\ref{eq:fkpkb}) one obtains 
succesively
\begin{equation}
-2(sT)' + \kappa (sR)' = 0 \;, \qquad s \kappa R= 2sT \; ,
\end{equation} 
wherefrom the analogue of the balance condition Eq.~(\ref{eq:balance})
follows
\begin{equation}
 \kappa R= 2T \; .
\label{eq:gs_balance}
\end{equation}

The link to the equations of motion Eqs.~(\ref{eq:eom}) is obtained by
taking the $k$-th moment of both sides of the Fokker-Planck
Eqs.~(\ref{eq:fkpk}) and using integration by parts for the terms
containing derivatives. The advantage of this approach is that not
only the equations of motions are recovered, but now the $k=0$ version of
the last one (Eq.~(\ref{eq:eomc})) can be obtained, and reads
\begin{equation} 
 g^2c_0- g^2v_0 + g^2 R_c(0) - \frac{-\kappa +\gamma +P}{2}t_0=0 \; ,   
\end{equation}
which amounts to extending the interpretation of $kc_{k-1}$ for $k=0$ by
the value of the GS function $R_c(s)$ at the origin. The process of
eliminating $c_k$ and $v_k$ in favour of $t_k$ can be repeated, but now
including the $k=0$ case. The balance relation Eq.~(\ref{eq:balance})
holds for $k=0$ too, owing to its GS version
Eq.~(\ref{eq:gs_balance}). As a result one recovers the three-term
recursion formula Eq.~(\ref{eq:recrel}) extended for $k=0$ as
\begin{equation} 
 A_0\,p_1 + B_0\, p_0 - \frac{P}{-\kappa+\gamma+P} \, R(0) = 0 \; .
\label{eq:recrel_0}
\end{equation}
This shows that $R(0)$ is the correct interpretation of $kp_{k-1}$
for $k=0$ \cite{fn-2}. In deriving the last equation we made use
of Eq.~(\ref{eq:fkpka}) which, for $k=0$ and $s=0 $, provides the
relationship between $R_c(0)$ and $R_v(0) = R(0)-R_c(0)$, allowing to
express $R_c(0)$ in terms of $R(0)$.     

Having established the generalization of the recursion relation to
$k=0$, it becomes clear that the ansatz contained in
Eq.~(\ref{eq:ansatz_n}), and which leads to the linear middle
behaviour Eq.~(\ref{eq:ansatz_cc}), is equivalent to the vanishing of
$R(0)$. We will therefore be concerned in what follows with the
circumstances in which this vanishing may take place.

One way to address this problem is to eliminate $R_c(s)$ and $R_v(s)$ from
Eqs.~(\ref{eq:fkpk}) in order to set up a differential equation for
$T(s)$ which, through Eq.~(\ref{eq:gs_balance}), provides a closed
equation for the purely photonic GS function $R(s)$. This essentially
parallels the procedure of obtaining the recursion relation, except that
we perform it at the level of the GS functions and not of their
moments. One obtains in this way a second-order differential equation
of the form
\begin{equation}
\begin{split}
(as^2+bs^3)R''(s) &+ (c+ds+es^2)R(s) \\
        &+(u+vs+ws^2)R(s) = 0 \; . 
\end{split}
\label{eq:R_ode}  
\end{equation}
Details about the derivation of this equation and the expression of
the coefficients in terms of the model parameters $\kappa$, $\gamma$
and the pump rate $P$ are given in the Appendix \ref{sec:R_ode}.  

The structure of this equation corresponds to $s=0$ being an
irregular singular point \cite{kmke}. Only one of the fundamental solutions   
behaves around $s=0$ as a power, $R(s) \sim s^{\sigma} $ and the
analysis leads to the result $\sigma=0$. The other fundamental solution
is highly singular $R(s) \sim exp(\lambda /s), \; \lambda > 0$
\cite{kmke}. These conclusions are parameter independent and show
that one cannot find a solution which vanishes at the
origin. 

This seems to contradict the previously mentioned numerical findings. The
paradox comes from the requirement $R(0)=0$, which is too
strong. In fact the numerical evidence hints only at $R(0)$ being
negligibly small. We show below that, indeed, the strong condition $R(0)=0$
holds only as a certain limiting case, which corresponds to a sharp
laser transition. Away from this limit the
transition gets smoother and the condition is expected to hold only as
an approximation. This is a familiar situation, also encountered in
connection with the many-emitter lasers \cite{rice-carm}. The
scaling limit procedure found in that context is based on 
an analysis performed at the level of the laser rate equation. In the
present case too, we first take guidance from the rate equation
approximation to find the appropriate limiting regime, which we later
analyze at the level of the exact solution.  

\section{The rate equation approximation}
\label{sec:rateq}    
 
The infinite hierarchy of the equations of motion  Eq.~(\ref{eq:eom})
can be interrupted by assuming that higher-order expectation values
can be approximately factorized into products of lower-order ones.
The factorization leads to the rate-equation approximation to the
full Liouville-von Neumann problem \cite{rice-carm}. In our case,
using
\begin{eqnarray}
 c_1 &=& \left < c^{\dagger}c \,  b^{\dagger}b \right> \approx 
             \left < c^{\dagger}c \right> \left < b^{\dagger}b \right>
             = c_0\, p_1 \; ,   \nonumber  \\
 v_1 &=& \left < v^{\dagger}v \,  b^{\dagger}b \right> \approx 
             \left < v^{\dagger}v \right> \left < b^{\dagger}b \right>
             = v_0\, p_1 \; , 
\label{eq:factor}    
\end{eqnarray}        
and limiting ourselves to the steady state, we are left with the
following closed set of equations for the quatities of interest
$n_c=c_0$, $n_v=v_0$ and $n=p_1$
\begin{eqnarray}
 n_c + n_v &=& 1 \; ,\nonumber \\
 -\gamma \,n_c +P\, n_v &=& \kappa \,n \; , \nonumber \\
 \Gamma \,n_c\,(n+1) - \Gamma \,n_v\,n &=&  \kappa\, n \; .
\label{eq:rate_cvn}
\end{eqnarray} 
Here $\Gamma = 4g^2/(\kappa+\gamma+P)$ is the rate of spontaneous
emission into the laser mode, and the physical interpretation of different
terms is obvious. Let us mention only the fact that the decrease of
$\Gamma$ with $P$ is a reflex of the dephasing effect of the pump,
which leads eventually to the quenching of photon output. 

Solving the first two equations for the level occupancies in terms of
the photon number leads to 
\begin{equation}
 n_c = \frac{P-\kappa\,n}{\gamma+P} \;, \qquad 
 n_v = \frac{\gamma+\kappa\,n}{\gamma+P} \; ,
\label{eq:fcv}
\end{equation}
which, introduced in the last one, gives rise to a quadratic equation
for $n$ 
\begin{eqnarray}
2\,(\kappa\,n)^2 + && \left[\kappa+\gamma-P +
   \frac{\kappa(\gamma+P)(\gamma+\kappa+P)}{4g^2} \right]\,(\kappa\,n) 
                                                   \nonumber\\
  && \hspace{3cm}-\, \kappa \,P = 0 \; ,
\label{eq:rate}
\end{eqnarray}
The two solutions have opposite signs and, of course, only the
positive one is physical. 

For the sake of the following discussion the above equation was
rewritten in terms of the rescaled photon number $\tilde n
=\kappa\,n$. In this form, in the limit $\kappa \to 0$ the free term
disappears and one of the roots becomes $\tilde n = 0$. 
This vanishing solution holds in the parameter domain in which the
other root is negative but, when positive, it is this second root that
is retained. Therefore an abrupt change of behaviour takes place when
the non-zero solution changes sign. As a function of the pump rate
the second root is a quadratic polynomial, initially increasing
but then decreasing, as predicted by the quenching effect. This
behaviour is preserved even in the  $\kappa \to 0$ limit, provided one
takes simultaneously $g \to 0$ in such a way that 
$\tilde g^2 = g^2/\kappa$ stays finite.  In this limit the solution
reads
\begin{equation}
  \tilde n = \frac{P-\gamma}{2}-\frac{(\gamma+P)^2}{8 \tilde g^2} = \nu(P)  
\label{eq:nu}
\end{equation}
when $\nu(P) \geqslant 0$ and  $\tilde n = 0$ otherwise. For the
upper-state population this leads to
\begin{equation}
 n_c = \begin{cases}
         \frac{P}{\gamma + P} & \text{if } \nu(P) <0 \\
         \frac{1}{2} + \frac{\gamma + P}{8 \tilde g^2} &\text{if }
         \nu(P) \geqslant 0
       \end{cases} 
\label{eq:nc_cases}
\end{equation}   
These solutions are shown in Fig.3. It is clear that the concave
dependence of $n_c$ on $P$ is interrupted by a linear middle
behaviour, in the interval defined by the roots of $\nu(P)$.  

\begin{figure*}[!htb]
\begin{center}
\resizebox{0.85\textwidth}{!}{%
 \includegraphics{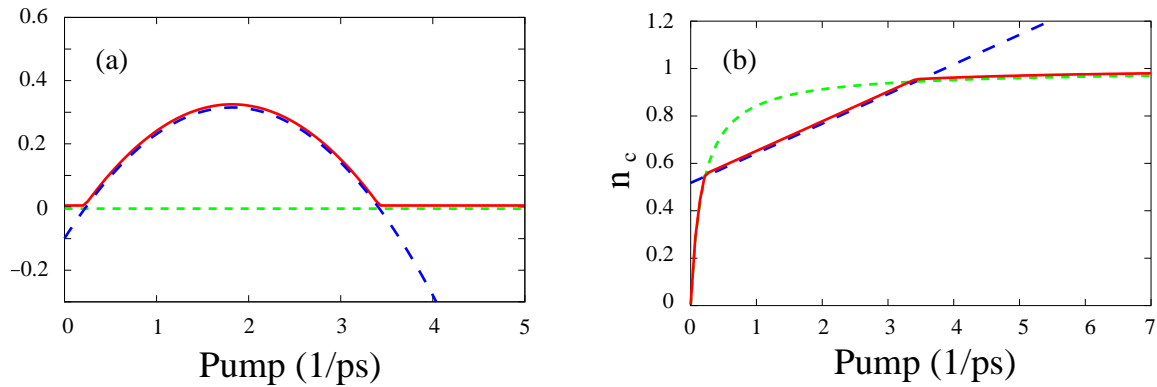}} 
\caption {\label{fig:Fig3}  (a) $\tilde n$ (full red line) is equal to
  $\nu(P)$ (blue dashed line) when the latter is positive and zero
  (green dashed line) otherwise. (b) the values of the upper level
  population $n_c$ for these cases, Eq.~(\ref{eq:nc_cases}), 
  showing a linear middle behaviour.    
}
\end{center}
\end{figure*} 

The situation described by the rate equation approximation suggests
that, in the above-defined scaling limit, the interval of positive
values for $\nu(P)$ coincides with the interval in which $R(0)=0$, and
in which the linear middle ansatz Eq.~(\ref{eq:ansatz_n}) holds. This
is indeed the case, as will be proven in Sec.~\ref{sec:scale}.

Several comments are in order. First, the appearance of the transition
is conditioned by existence of real, positive roots of the equation
$\nu(P)=0$. It is easy to see that real roots exist if $\tilde g^2
\geqslant 2 \gamma$, or $g^2 \geqslant 2\kappa \gamma$, and then they are
both positive. The two types of behaviour discussed in Fig.1 are
distinguished by whether this condition is fulfilled or not.    

Second, the scaling procedure defined here $\kappa\to 0$, $g \to 0$,
$g^2 / \kappa$ finite, corresponds entirely to the limiting regime
($\kappa \to 0, \beta \to 0$, $\beta / \kappa$ finite) considered by Rice and
Carmichael \cite{rice-carm}, because the $\beta$-factor is proportional to
$g^2$. Moreover, the Jaynes-Cummings coupling constant appears as
an explicit parameter in all laser models and therefore the
scaling limit formulated in terms of $g$ is not limited to the
two-level atom model (see Appendix \ref{sec:rand_inj}).  

Finally, since $\tilde n$ is the rate of photon emission out of the
cavity, its nonvanishing value as $\kappa$ goes to zero expresses the
appearance of a finite laser output even as we get arbitrarily close
to a perfect cavity, by a corresponding increase of the photon generation
inside. This is physically interpreted as ``an explosion of stimulated
emission'' \cite{rice-carm}, typical for the lasing regime.
  
The rate equation approach provides only partial informations about
the photon system. The full photon statistics is contained in
the GS function $R(s)$, whose scaling behaviour is addressed in the
next Section.

\section{The scaling limit of $R(s)$ }
\label{sec:scale}   

As $n \to \infty $, the moments of $R(s)$ cease to exist, showing that 
the GS function tends to move its weight on larger and larger
$s$ values. In this situation, meaningful quantities are the rescaled
ones, like
\begin{equation}
   \tilde n = \tilde p_1 = \int_0^\infty \kappa\, R(s)\, s\, \text{d}s
   \; .
\end{equation}
In the rescaled variable $t = \kappa s$, one has 
\begin{equation}
  \tilde p_1 = \int_0^\infty \tilde R(t)\, t\, \text{d}t \; ,
\end{equation} 
where the rescaled GS function $\tilde R(t)$ is defined by
\begin{equation}
   \tilde R(t) = \frac{1}{\kappa}\, R \left(\frac{t}{\kappa}\right) \; .
\label{eq:R_t}
\end{equation}  
Similarly, all the higher moments $p_k$ are rescaled as $\tilde p_k =
\kappa ^k p_k$ and become corresponding moments of $\tilde R(t)$. 
\begin{equation}
 \tilde p_k = \int_0^\infty \tilde R(t)\, t^k\, \text{d}t \; .
\label{eq:tdpk_mom}
\end{equation}

The task now is to establish the equation obeyed by $\tilde R(t)$ in
the scaling limit and to study the behaviour of its solution around
the origin. Intuitively, according to Eq.~(\ref{eq:R_t}), the graph of
$\tilde R(t)$ is obtained from that of $R(s)$ by compressing the
latter by a factor of $1/\kappa$ along the abscissa and expanding it
by the same factor along the ordinate. This would bring the rescaled
function to $\delta(t)$, in the limit $\kappa \to 0$, were it not for
the opposite tendency of $R(s)$ to move away from the origin, as
discussed above. The net result of these competing trends is analyzed
below.

To this end, one translates Eq.~(\ref{eq:R_ode}) obeyed by $R(s)$ into
an equation for $\tilde R(t)$, retaining in the resulting coefficients
of the latter only the dominant terms in the scaling parameter
$\kappa$. One obtains \cite{fn-3}
\begin{equation}
\begin{split}
\frac{t^2}{4\tilde g^2}\, \kappa^2 \, \tilde R''
     & - \left[ \left(3\, \frac{\gamma+P}{8\tilde g^2} +1 \right) t 
        -\frac {1}{2}\, P \right] \, \kappa \, \tilde R' \\
     & + \left [t - \nu(P) \right] \, \tilde R = 0  \; .
\end{split}
\label{eq:R_t_ode}
\end{equation}  
The appearance of the small parameter $\kappa$ with the derivatives
suggests a WKB approach to the $\kappa \to 0$ asymptotics of the
solution. In other words one searches the solution, up to a
normalization factor, in the form
\begin{equation} 
\tilde R(t) = e^{-\frac{1}{\kappa} \varphi(t)} \; ,  
\label{eq:wkb} 
\end{equation}      
in which $\varphi(t)$ is taken in the leading, zeroth order in
$\kappa$. It is clear that when $\kappa$ gets smaller, the value of 
$\tilde R(t)$ around the minimum of $\varphi(t)$ is greatly enhanced,
at the expense of the values at other points which, in the view of
normalization, become negligible. In the limit one obtains a
$\delta$-function concentrated at the minimum of $\varphi(t)$.   

The equation obeyed by  $\varphi(t)$ in the leading order has the form
of a quadratic equation for its derivative
\begin{equation} 
\frac{t^2}{4\tilde g^2}\, (\varphi')^2
      + \left[ \left(3\, \frac{\gamma+P}{8\tilde g^2} +1 \right) t 
        -\frac {1}{2}\, P \right] \, \varphi'
      + \left [t - \nu(P) \right] = 0  \; .
\label{eq:d_phi}
\end{equation}   
Around $t=0$ one of the roots behaves like $\varphi'\sim 2\tilde g^2
P/t^2$, i.e.  $\varphi \sim - 2\tilde g^2 P/t $  which, by
Eq.~(\ref{eq:wkb}), leads to the singular solution mentioned in 
Section~\ref{sec:GS}. The regular solution comes from the other root,
for which $\varphi'(0) = -2\nu(P)/P$. 

Two cases arise, depending on the sign of $\nu(P)$: (i) As long as
$\nu(P)$ is negative one has $\varphi'(0) >0$ and $t=0$ is a minimum
for $\varphi(t)$. Then, according to the above discussion, $\tilde
R(t)$ tends to $\delta(t)$ in the scaling limit. 
(ii) When $\nu(P)$ becomes positive,
$\varphi'$ starts at $t=0$ with negative values and crosses the
abscissa at $t=\nu$. Then $\varphi(t)$ has a local maximum at the
origin and therefore the values of $\tilde R(0)$ become vanishingly
small in the limit $\kappa \to 0$. This is precisely the requirement
for the linear middle behaviour to hold. The values of $\tilde R(t)$
are now concentrated at the point of minimum $t=\nu(P)$. In other
words $\tilde R(t) \to \delta(t-\nu)$. 
    
In the first case, according to Eq.~(\ref{eq:tdpk_mom}), all rescaled 
expectation values are vanishing in the scaling limit, except $\tilde
p_0 = 1$,  while in the
second regime they become $\tilde p_k = \nu^k$. This exponential 
behaviour of the normal-ordered averages $\left < b^{\dagger \, k} b^k
\right >$ is equivalent to a Poissonian photon number statistics. The
change between the two regimes occurs when $\nu(P)$ changes sign, from
negative to positive and back, in agreement with the result suggested
by the rate equation approximation. 

In terms of the phase-space complex argument $\alpha$, the rescaled 
GS function $\tilde R(t)$ changes during the transition from being
concentrated around the origin into a infinitely thin ring-like distribution
having a radius of $\sqrt{\nu(P)}$. For the unscaled GS
function $R(s)$, this corresponds to a ring distribution 
whose radius grows in the scaling limit faster than its
thickness \cite{fn-4}. This is how its value at the origin becomes negligibly
small, in accordance with the discussion in
Section~\ref{sec:cont_frac} regarding why and when is the ansatz of
Eq.~(\ref{eq:ansatz_n}) valid.   

\section{Conclusions and discussion}
\label{sec:conc} 

The numerical solution of the problem of a two-level system in
interaction with a cavity mode shows a rather sudden change in the
behaviour of the steady-state data as a function of pumping. Using  
an analytic approach, we reproduce the numerical results and discuss
this regime change. 

One of the main tools used was the Glauber-Sudarshan representation,
which was shown to provide the natural extension,
Eq.~(\ref{eq:recrel_0}), to the recursion relation
Eq.~(\ref{eq:recrel}). In this perspective it becomes clear that the
transition observed is related to the vanishing of the GS
quasi-distribution function $R(s)$ at the origin.  

The laser transition is often represented as the tendency of the GS
function to move away from the origin, and this is exactly what takes
place in this case too. More precisely, it was shown that in the limit
$\kappa \to 0$, $g \to 0$ so that $g^2/\kappa=const$, the GS function
becomes in the lasing regime a $\delta$-like distribution concentrated
on a ring with a pump-dependent radius. In terms of photon number
statistics this means that the distribution becomes Poissonian. 

It is instructive to see the action of the scaling limit directly on
the recursion relation Eq.~(\ref{eq:recrel}). Rewritten for the rescaled
quantities  $\tilde p_k= \kappa^k p_k$ its coefficients are changed
respectively into $\tilde A_k = A_k/\kappa$, $\tilde B_k = B_k$ and
 $\tilde C_k = \kappa \,C_k$. In the scaling limit these coefficients
become $k$-independent
\begin{align} 
&\tilde A_k \to \frac{2}{P+\gamma} \, , \nonumber \\ 
&\tilde B_k \to \frac{-P+\gamma}{P+\gamma} +\frac{P+\gamma}{8\tilde
    g^2}  \, , \nonumber \\
& \tilde C_k \to 0 \, , 
\label{eq:lim_ABC}
\end{align}
with the ratio $\tilde A_k/ \tilde B_k$ converging to $\nu(P)$. Then
the recursion relation becomes simply  $\tilde p_{k+1}= \nu \,
\tilde p_k$ for $k \geqslant 1$. For the values of the pump for which $\nu$ is
negative only the trivial solution $\tilde p_k =0$ for all $k
\geqslant 1$ is allowed, to avoid the appearance of negative
expectation values. Conversely, when $\nu(P)$ is positive a nontrivial
solution becomes  possible. Moreover, according to the above
discussion, in this case $\tilde R(0)=0$ and then the additional
relation $\tilde p_1= \nu \, \tilde p_0 = \nu$, deduced from
Eq.~(\ref{eq:recrel_0}) also holds.  
Then the solution is $\tilde p_k= \nu^k$, in agreement with the 
statement that one has $\tilde R(t) \to \delta(t-\nu)$ in the scaling
limit and the photon statistics becomes purely Poissonian. 

Examining the convergence in Eq.~(\ref{eq:lim_ABC}) it is clear that as
$\kappa$ becomes smaller and smaller it is the low-$k$ coefficients
that get close to their limit first. This is because the limit
requires the product $k \kappa$ to be small. Therefore the
statistics becomes accurately Poissonian for the low-index $p_k$
first. As $\kappa$ further approaches zero the list of
nearly-Poissonian expectation values grows longer and longer and only
in the limit it extends to the whole sequence. In view of these facts
$g^{(2)}(0)$ gets close to unity quite early in the process, before
the statistics becoming entirely Poissonian. As a symptom of
a truly coherent light, the condition $g^{(2)}(0)=1$ may be, in this
sense, somewhat premature.

The scaling limit corresponds to a weak coupling of the cavity both to
its photon source and to its drain. Both coupling parameters involved
are present in all laser models, so that the limiting procedure can
be tested, in principle, on any of them, and we have shown its
validity of the Scully-Lamb model too. As a procedure which is
successful in two unrelated cases, the scaling limit has a good chance
of having a wider applicability.


\appendix
\section{Differential equation for the Glauber-Sudarshan function $R(s)$ }
\label{sec:R_ode}    

The second order differential equation obeyed by $R(s)$ is deduced
from Eqs.~(\ref{eq:fkpk}) as follows. We note first that $R_v$ and $T$
can be eliminated in favour of $R$ and $R_c$, the first by
$R_v=R-R_c$ and the second using the balance condition
Eq.~(\ref{eq:gs_balance}). As a second step one solves for $R_c$ and
$R'_c$ in favour of $R$ and then identifies the derivative of the first
with the second. This way Eq.~(\ref{eq:R_ode}) is obtained, with the
following coefficients              
\begin{align}
a &=\kappa \frac{\gamma-\kappa+P}{g^2} \; , \nonumber \\
b &= -2 \frac{\kappa^2}{g^2} \; , \nonumber \\
c &=  \frac{2P(\gamma - \kappa + P)}{\kappa^2} \; , \nonumber \\
d &= - \left[4\frac{P}{\kappa} +\frac{\gamma -\kappa +P}{\kappa}
            \left(4 + 3 \kappa \frac{\gamma - 3 \kappa
              +P}{2g^2}\right) \right] \; , \nonumber \\
e &= 8 + \kappa\frac{3\gamma - 7\kappa +3P}{g^2} \; , \nonumber \\
u &= -\frac{4P(\gamma -2\kappa+P)}{\kappa^2} \nonumber \\ 
  & \hspace{-0.2cm}+ \frac{\gamma -\kappa
  +P}{\kappa}\left[2\frac{\gamma -\kappa+P}{\kappa}
  +\frac{(\gamma-2\kappa+P)(\gamma-3\kappa+P)}{2g^2} -4\right] , \nonumber \\
v &= 8\frac{P}{\kappa} - \left( \gamma -\kappa +P \right)\frac{\gamma
  -3\kappa +P}{g^2} \; , \nonumber \\
w &= -8 \; . 
\end{align} 

\section{The random injection model}
\label{sec:rand_inj} 

In this Appendix we apply the scaling limit to the random injection
model, in order to show that in this case too a sharp transition to
Poissonian distribution takes place, in the same parameter region as 
given in the literature.

In the Scully-Lamb laser model \cite{scl-lamb,sten, walls}
the cavity mode is fed by random injection of
inverted two-level atoms, which exchange energy with the mode through
Jaynes-Cummings interaction. The time each atom spends inside the
cavity is a random variable too. The losses are simulated by a similar
random injection of atoms, this time in their lower state. One obtains
for the time evolution obeyed by the photonic density matrix
\begin{equation}
\begin{split}
\frac{\partial}{\partial t} \rho_{n,n} =&
       \frac{A\,n}{1+\frac{B}{A}\,n}\, \rho_{n-1.n-1} 
      -\frac{A\,(n+1)}{1+\frac{B}{A}\,(n+1)}\, \rho_{n.n} \\
      &-C\,n\,\rho_{n.n} + C\,(n+1)\,\rho_{n+1.n+1} \; .
\end{split}
\label{eq:rand_inj}  
\end{equation}
We keep here the notation commonly found in the literature
\cite{scl-lamb,sten, orsz} for the resulting parameters.  
For $B=0$ one would obtain the usual competition between the loss
factor $C$ and the gain (or amplification) factor $A$. This is
known to lead to an unstable behaviour for $A>C$. The role of the
saturation parameter $B>0$ is to limit the photon output and to
stabilize the solution. A common procedure in the analysis of the
problem is to expand the denominators of Eq.~(\ref{eq:rand_inj}) up to
the first order in $B$, but we will not resort to this approximation. 

In terms of the Jaynes-Cummings coupling constant one has $A$ and $C$
proportional to $g^2$ and $B$ to $g^4$. Since the cavity loss rate
$C$ is the same as the parameter $\kappa$ used throughout the paper,
the limit $\kappa \to 0$  naturally implies that $g^2 \to 0$ too,
with their ratio remaining constant.    

It is easily seen that the steady-state solution of
Eq.~(\ref{eq:rand_inj}) obeys 
\begin{equation}
  \rho_{n+1,n+1} = \frac{\mu}{\beta+1+n}\, \rho_{n,n} \; ,
\label{eq:rec_rho}
\end{equation} 
where the notation $\mu = A^2/BC$ and $\beta = A/B$ has been
used. This leads to
\begin{equation}
 \rho_{n,n} =  \frac{\mu^n}{(\beta+1)(\beta+2)\ldots (\beta+n)}\,
 \rho_{0,0} \; .
\label{eq:rho_n}
\end{equation} 
As it was early recognized \cite{scl-lamb}, the generating
function of $\rho_{n,n}$ is given, up to normalization, by Kummer's
confluent hypergeometric function
\begin{equation} 
\begin{split}
\rho(x)& = \sum_n\, \rho_{n,n}\, x^n  
         = \rho_{0,0}  \sum_n\ \,
\frac{(\mu\,x)^n}{(\beta+1)(\beta+2)\ldots (\beta+n)} \\
       & =\rho_{0,0} \;_1F_1(1,\beta+1,\mu x) \; . 
\end{split}
\label{eq:rho_x}
\end{equation}  
The value of $\rho_{0,0}$ is determined by the normalization condition
$\rho(x)=1$. The photon number distribution $\rho_{n,n}$ is connected
to the GS function $R(s)$ by Eq.~(\ref{eq:rho_n_gs}) 
which implies, with $y=1-x$ and $z=y/\kappa$
\begin{equation}    
\begin{split}   
  \rho(x) = \int_0^{\infty} R(s)\,e^{-ys} \text{d}\,s 
          &= \sum_{k=0}^{\infty} \frac{(-1)^k}{k!}\,y^k\,p_k \\ 
          &= \sum_{k=0}^{\infty} \frac{(-1)^k}{k!}\,z^k\,\tilde p_k  \; .        
\end{split}
\label{eq:rho_x_gs} 
\end{equation} 
This shows that $\rho(x)$ is also the generating function for
the expectation values $p_k$  and of the rescaled ones $\tilde p_k$
when seen as a function of $y$ and $z$, respectively. Therefore 
$\rho(x)$ is the appropriate tool for analyzing the scaling behaviour
of these quantities. According to their definition, both $\mu$ and $\beta$
scale like $1/g^2$ so they go to infinity as $\kappa$ goes to zero, in
such a way that $\tilde \mu = \kappa \mu$ and $\tilde \beta = \kappa
\beta $ remain finite.   

At this point one can make use of the integral representation of
Kummer's function \cite{ab-steg}
\begin{equation} 
_1F_1(1,\beta +1, \mu x) = \beta \int_0^1 e^{\mu x u} \,(1-u)^{\beta-1}
  \text{d} \, u \; ,
\label{eq:int_rep} 
\end{equation}  
to express the generation function for $\tilde p_k$ as
\begin{equation}
\tilde p(z) = \rho(1-\kappa z) = \frac{I(z)}{I(0)} \; ,
\label{eq:norm}
\end{equation}
with the notation 
\begin{equation} 
I(z) = \int_0^1  e^{(1-\kappa z) u \tilde \mu/\kappa} \, (1-u)^{\tilde
  \beta /\kappa  -1} \text{d} u \; ,  
\end{equation}
and division by $I(0)$ is necessary to ensure the normalization
condition $\tilde p(0)= \rho(1)=1$.  
Rearranging $I(z)$ as
\begin{equation} 
I(z) = \int_0^1  e^{- z u \tilde \mu} \, e^{[\tilde \mu u
  +\tilde \beta \ln(1-u)]/\kappa}\, (1-u)^{-1} \text{d} u \; ,  
\end{equation}
one brings it into a form familiar from the method of steepest
descent. The $\kappa \to 0$ asymptotics of $I(z)$ is controlled by the 
maximum of $\varphi(u)=\tilde \mu u + \tilde \beta \ln (1-u)$ in the
interval of integration, since the contribution of the integrand
around the maximum point becomes overwhelmingly dominant. Depending on
the position $u_0$ of the saddle point, defined by $\varphi'(u_0)=0$,
i.e.
\begin{equation}
u_0 = 1-\frac{\tilde \beta}{\tilde \mu } = 1-\frac{\beta}{\mu } 
    = 1-\frac{C}{A} \; ,
\end{equation} 
one has two regimes. (i) When $A<C$, $u_0$ is negative and the maximum of
$\varphi(u)$ on $[0,1]$ is found at $u=0$. (ii) When $A>C$ the saddle
point enters 
the integration interval and the maximum of $\varphi(u)$ is reached
for $u=u_0$. As a result, and given the normalization condition
Eq.~(\ref{eq:norm}), one obtains
\begin{equation}
\tilde p(z) = \begin{cases}
               1           & \text{if } A \leqslant C \, ,\\
               e^{-z\mu u_0}  & \text{if } A>C \; .
              \end{cases} 
\end{equation} 
In the first case all $\tilde p_k=0$ with the exception of $\tilde
p_0=1$, while in the second one obtains $\tilde p_k=\nu^k$ with
\begin{equation}
\nu=\tilde \mu \,u_o =\frac{A(A-C)}{BC} \; .
\end{equation}
One finds again the same scenario as in Section~\ref{sec:scale}. The
rescaled expectation values $\tilde p_k$ obey an exponential
$k$-dependence, with a base that changes from zero to a positive
$\nu$ at the threshold point. The expression of $\nu$ found by
the present procedure is the same as given in the literature
\cite{scl-lamb, sten, orsz}.


\begin{thebibliography}{99}
\bibitem{fn-1} Pringsheim's convergence criterion \cite{cont-fr} is
obeyed for k sufficiently large
\bibitem{fn-2} The result is natural in the light of the following
heuristic argument: with $k$ in Eq.~(\ref{eq:pk_mom}) as a continuous
positive parameter, one has 
$kp_{k-1}= \int_0^{\infty}R(s)(s^k)'{\text d}s$
which, after an integration by parts followed by the limit $k \to
0$, leads to $R(0)$.
\bibitem{fn-3} A common factor $t-(\gamma+P)/2$
appears in the process and is reduced. Therefore the coefficients have
a lower degree than in Eq.~(\ref{eq:R_ode}).
\end{thebibliography}
\end{document}